# Computation of Electrostatic Properties of 3D MEMS structures

N.Majumdar, S.Mukhopadhyay

*Saha Institute of Nuclear Physics,*

*1/AF, Sector 1, Bidhannagar, Kolkata 700064, West Bengal, India*

## INTRODUCTION

Micro-Electro-Mechanical Systems (MEMS) normally have fixed or moving structures with cross-sections of the order of microns (μ*m*) and lengths of the order of tens or hundreds of microns. These structures are often plates or array of thin beams which, owing to their smallness, can be moved or deflected easily through the application of low voltages. These structures are widely used in microjets, microspeakers, electrostatic actuators etc. Since electrostatic forces play a very major role in maneuvering these devices, a thorough understanding of the electrostatic properties of these structures is of critical importance, especially in the design phase of MEMS.

In many cases, the electrostatic analysis of MEMS is carried out using boundary element method (BEM), while the structural analysis is carried out using finite element method (FEM) [1]. In this paper, we focus on accurate electrostatic analysis of MEMS using BEM. In particular, we consider the problem of computing the charge distribution and capacitance of thin conducting plates relevant to the numerical simulation of MEMS. Computing the resultant charge due to the two surfaces of a plate under the usual assumption of vanishing thickness of the plate is not an acceptable approach for these structures. This is so because the electrostatic force acting at any point on these surfaces depends on the square of the charge density at that point. As a result, the standard BEM does not work satisfactorily and several modified BEM have been developed, such as the enhanced BEM and the gradient boundary integral equation (BIE) technique [2]. The former is suitable for moderately thick plates, while the latter is suitable for very thin plates, $h/L <= 10^{-3}$, where *h* is the thickness and *L* is the length of a side of a square plate.

In this work, we present a BEM solver which is equally accurate for plates of any thickness and applicable for any structure for which electrostatic properties need to be computed, including those relevant for MEMS.

The reason behind the accuracy of the solver is the fact that it uses closed-form analytic expressions that are valid seamlessly throughout the physical domain for computing the influence coefficients. Thus, it is possible to avoid one of the most serious approximations of the BEM, namely, the assumption that the effect of a charge distributed over a boundary element can be approximated by charge located at the centroid of the element. Here, we concentrate on the charge density and the capacitance of the structure, both of which are very important for the design of efficient MEMS. Comparison with other results available in the literature seems to indicate that the present results are more accurate than the existing ones.

## THEORY

The potential at a point P(X,Y,Z) in space due to an uniform charge distributed on an element on the XZ plane can be obtained in closed-form. This analytic expression is exact and valid for the complete physical domain [3], [4]. Since no approximation needs to be made during the derivation of this expression, it is equally applicable in cases where surfaces (such as upper and lower surfaces of plates) are far apart or very near to each other. These expressions have been used to develop a new BEM solver which has been applied to solve the problem of finding out the charge density on the four surfaces of two plates of $h/L <= 10^{-6}$ and various values of plate separation, $d$, as shown in Figure 1.

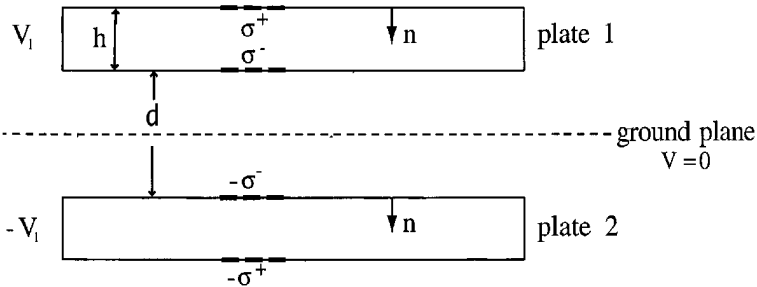

*Figure 1 Geometry of the problem*

## RESULTS

In Figure 2, we present the variation of the normalized capacitance ($Cd/A\epsilon$) of the system with changing $d/L$. Our results have been compared with those from [2] and [5]. It can be seen that the present results fall in between those of the other two, being very close to [2]. From this figure we can conclude that our results follow the general trend available till date.

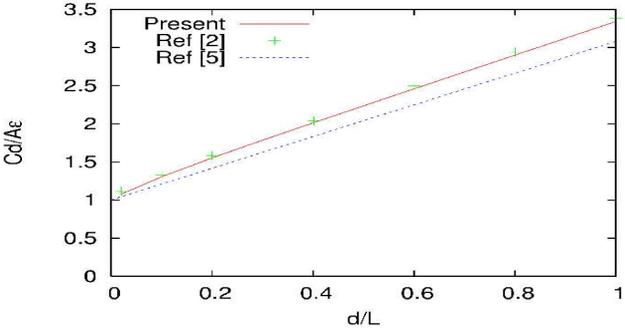

*Figure 2Normalized capacitance vs. normalized plate separation*

In Figure 3, we present the charge density along a midline on the upper and lower surfaces of the upper plate of Figure 1 and compare our values with those from [2]. In this case, the h/L = $10^{-3}$ while *d/L* = 0.2. The values agree quite well except that we have not found any 'bump' towards the end of the plate. In fact, such bumps could be observed only when we used very coarse discretization and is probably an artifact, as has been remarked in [2]. The fact that the new solver computes charge density without the 'bumps' have led us to believe that the present results are more accurate than those available. Since these artifacts are likely to over-predict the total charge on the plate, the resulting normalized capacitance is possibly also over-predicted in [2].

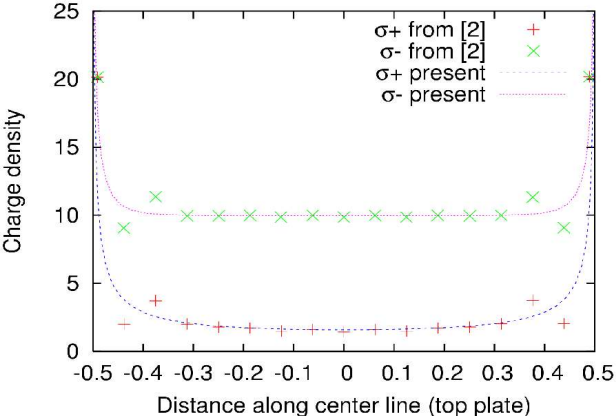

*Figure 3Comparison of charge density on the top plate*

## CONCLUSIONS

A recently developed BEM solver has been used to compute the electrostatic properties of structures relevant to MEMS. The solver is accurate and applicable to a wide range of problems because the influence coefficients are calculated using exact analytic expressions. The obtained results have been compared with other available results. The comparison indicated that the presented ones are possibly more accurate than the available results.

## ACKNOWLEDGEMENT

The authors are grateful for the support extended by Prof. Bikash Sinha, Director, SINP and Prof. Sudeb Bhattacharya, HOD, NAP, SINP.